\documentclass[11pt,twoside]{article}
\usepackage{./asp2014}

\aspSuppressVolSlug
\resetcounters

\begin{document}

\title{
Examining magnetospheric accretion in Herbig Ae/Be stars through near-infrared
spectroscopic signatures
}
\author{
Markus~Sch\"oller,$^1$
Swetlana~Hubrig,$^2$
Mikhail~A.~Pogodin,$^3$
Silva~P.~J\"arvinen,$^2$
Natalia~A.~Drake,$^{4,5,6}$
and J.~Andres~Cahuasqui$^7$
\affil{$^1$European Southern Observatory, Karl-Schwarzschild-Str.~2, 85748~Garching, Germany; \email{mschoell@eso.org}}
\affil{$^2$Leibniz-Institut f\"ur Astrophysik Potsdam (AIP), An der Sternwarte~16, 14482~Potsdam, Germany}
\affil{$^3$Central Astronomical Observatory at Pulkovo, Pulkovskoye chaussee~65, 196140~Saint~Petersburg, Russia}
\affil{$^4$Laboratory of Observational Astrophysics, Saint Petersburg State University, Universitetsky~pr.~28, 198504~Saint~Petersburg, Russia}
\affil{$^5$Observat\'orio Nacional/MCTIC, Rua General Jos\'e Cristino~77, CEP~20921-400, Rio~de~Janeiro, RJ, Brazil}
\affil{$^6$Laborat\'orio Nacional de Astrof\'{\i}sica/MCTIC, Rua Estados Unidos 154, 37504-364 Itajub\'a, MG, Brazil}
\affil{$^7$I.~Physikalisches Institut, Universit\"at zu K\"oln, Z\"ulpicher Str.~77, 50937~K\"oln, Germany}
}

\paperauthor{Markus~Schoeller}{mschoell@eso.org}{0000-0002-5379-1286}{European Southern Observatory}{}{Garching}{}{85748}{Germany}
\paperauthor{Swetlana~Hubrig}{shubrig@aip.de}{0000-0003-0153-359X}{Leibniz-Institut f\"ur Astrophysik Potsdam (AIP)}{}{Potsdam}{}{14482}{Germany}
\paperauthor{Mikhail~A.~Pogodin}{pogodin@gaoran.ru}{}{Central Astronomical Observatory at Pulkovo}{}{Saint Petersburg}{}{196140}{Russia}
\paperauthor{Silva~P.~Jarvinen}{sjarvinen@aip.de}{}{Leibniz-Institut f\"ur Astrophysik Potsdam (AIP)}{}{Potsdam}{}{14482}{Germany}
\paperauthor{Natalia~A.~Drake}{natalia.drake.2008@gmail.com}{0000-0003-4842-8834}{Saint Petersburg State University}{}{Saint Petersburg}{}{198504}{Russia}
\paperauthor{J.~Andres~Cahuasqui}{cahuasqui@ph1.uni-koeln.de}{}{Universit\"at zu K\"oln}{I.~Physikalisches Institut}{K\"oln}{}{50937}{Germany}

\begin{abstract}
Models of magnetically driven accretion and outflows reproduce many observational properties of T\,Tauri stars.
For the more massive Herbig~Ae/Be stars, the corresponding picture is not well established.
Nonetheless, it is expected that accretion flows in pre-main-sequence stars are guided
from the circumstellar disk to stellar regions of high latitude along the magnetic field lines inside a magnetosphere.
Using near-infrared multi-epoch spectroscopic data obtained with ISAAC, CRIRES, and X-shooter on the VLT,
we examined magnetospheric accretion in the two Herbig~Ae stars HD\,101412 and HD\,104237.
Spectroscopic signatures in He\,{\sc i}~10\,830 and Pa$\gamma$,
two near-infrared lines that are formed in a Herbig star's accretion region, show temporal modulation in both objects.
For HD\,101412, this modulation is governed by its rotation period, which we could recover from the data.
We could show that our spectroscopic observations can be explained within the magnetic geometry
that we established earlier from magnetic field measurements.
For HD\,104237, we struggled to clearly identify a rotation period.
We intend to apply this method to a larger sample of Herbig Ae/Be stars
to learn more about their rotation properties and the accretion mechanisms at work.
\end{abstract}

\section{Introduction}

Herbig~Ae/Be stars (HAeBes) are predecessors of main-sequence stars in the mass range 2--10\,$M_{\odot}$. 
They show clear signatures of surrounding disks, as evidenced by a strong infrared excess, and are actively accreting material.
The phase between protostar and main-sequence object is a key stage for planet formation:
dusty disks provide the material needed for the formation of planets. 
In Herbig~Ae stars, observations suggest a close parallel to T\,Tauri stars,
i.e.\ the stellar magnetic field truncates the accretion disk at a few stellar radii
and gas accretes along magnetic field lines from the protoplanetary disk to the star
(magnetospheric accretion (MA); e.g.\ \citealt{Muzerolle2004}).
In Herbig~Be stars, it is presumed that the accretion flow is not disrupted by the magnetic field.

Before 2004, the only magnetic field detection had been reported for the optically brightest ($m_{\rm V}=6.5$)
Herbig~Ae star HD\,104237 \citep{Donati1997}, but no further publication confirming this detection existed.
Consecutive studies 
reported the discovery of magnetic fields in seven other HAeBes
(\citealt{Wade2005,Wade2007,Catala2007,Hubrig2004, Hubrig2006, Hubrig2007}). 
\citet{Alecian2008} reported eight magnetic HAeBes in a sample of 128~objects.
Later on, a study of 21~HAeBes with FORS\,1/2 revealed the presence of magnetic fields in six additional stars \citep{Hubrig2009}.
Further studies involved the outbursting binary Z\,CMa \citep{Szeifert2010},
the Herbig~Ae star HD\,101412 with resolved magnetically split lines \citep{Hubrig2010},
HD\,31648 \citep{Hubrig2011},
PDS\,2 \citep{Hubrig2015}, and
the two systems AK\,Sco and HD\,95881 \citep{Jarvinen2018}.

\section{Magnetospheric accretion in Herbig Ae/Be stars}

While MA is well established as the accretion scenario for
T\,Tauri stars, it is not so clear if this also holds for the HAeBes,
mainly driven by the facts that not many magnetic HAeBes have been found and that
their magnetic fields are typically an order of magnitude weaker.

\citet{Wade2007} used the equations put forward by \citet{JohnsKrull1999}
for T\,Tauri stars to estimate the magnetic field strength needed
to support MA in HAeBes.
Assuming
$v\,\sin\,i = 115$\,km\,s$^{-1}$,
$R_* = 2R_{\odot}$,
$i = 90^{\circ}$ (leading to $P_{\rm rot} = 1$\,d),
$M_* = 2M_{\odot}$,
$\dot M_{\rm acc} = 10^{-8}M_{\odot}$\,yr$^{-1}$,
which are canonical values for model parameters, one would need
$B_{\rm d} = 500$\,G for the models by \citet{Koenigl1991} and \citet{Shu1994} and
$B_{\rm d} = 100$\,G for the model by \citet{CollierCameronCampbell1993}.
The required field strengths increase with mass, period, and mass accretion rate, and decrease with radius.

According to \citet{Alecian2014},
the magnetic properties of A and B-type stars were shaped before the HAeBe evolutionary phase.
Using pre-main-sequence evolutionary tracks calculated with the CESAM code \citep{Morel1997},
the authors concluded that even stars above 3\,$M_\odot$ undergo a purely convective phase
before reaching the birthline.
Based on all available measurements summarized by \citet{Hubrig2015},
we can consider that it is reasonable to assume that the weak magnetic fields detected in a number of HAeBes are
leftovers of the magnetic fields generated by dynamos during these convective phases.

\citet{CauleyJohnsKrull2014}
studied the He~{\sc i}~10\,830 morphology in a sample of 56~HAeBes.
They suggested that early Herbig~Be stars do not accrete material from their inner disks
in the same manner as T\,Tauri stars,
while late Herbig~Be and Herbig~Ae stars show evidence for MA.
Furthermore, they proposed more compact magnetospheres in HAeBes compared to T\,Tauri stars.
Further, \citet{Ababakr2017} found that
42 of 56~HAeBes in their sample show a polarization change over the H$\alpha$ line,
which is attributed to a small scale asymmetry due to the surrounding disk.
The behavior of Herbig~Ae stars was found to be similar to T\,Tauri stars,
while Herbig~Be stars earlier than B7/B8 show a line depolarization.

Interferometric searches for magnetospheres in the near-infrared did not come to a final conclusion.
\citet{Kraus2008} observed five HAeBes with AMBER on the VLTI
and found only in HD\,98922 -- then not known to be magnetic --
a Br$\gamma$ line-emitting region compact enough to be compatible with a magnetosphere.
For the other four sources in the study, including the magnetic HD\,104237,
they found larger sizes consistent with an extended stellar wind or a disk-wind.

Studying MA in HD\,58647,
\citet{Kurosawa2016} modeled continuum and Br$\gamma$ AMBER data
and found that a disk wind plus a small magnetosphere explain all measurables.
J\"arvinen et al.\ (in prep.) found a change of magnetic field polarity on a timescale of days in this star.

Looking at observations obtained with the CHARA array in H$\alpha$, there is a wide spread
of results.
\citet{Perraut2016} found evidence for a disk-wind coming from 0.3\,au and for a magnetosphere in AB\,Aur,
\citet{Benisty2013} saw a disk-wind on a scale of 0.2--0.6\,au in MWC\,361, and
\citet{Mendigutia2017} determined a size of 15\,R$_*$ in HD\,179218,
which is likely due to MA,
and a size of 16\,R$_*$ in HD\,141569, for which MA is impossible due to the high rotational velocity of this star.
Overall, it is clear that a wider variety of scenarios
is needed to explain the H$\alpha$ emission:
coming from compact or extended sources, from a disk, from the accretion flows, and from winds.

\section{HD\,101412}

\begin{figure}
\centering
\includegraphics[width=0.36\columnwidth]{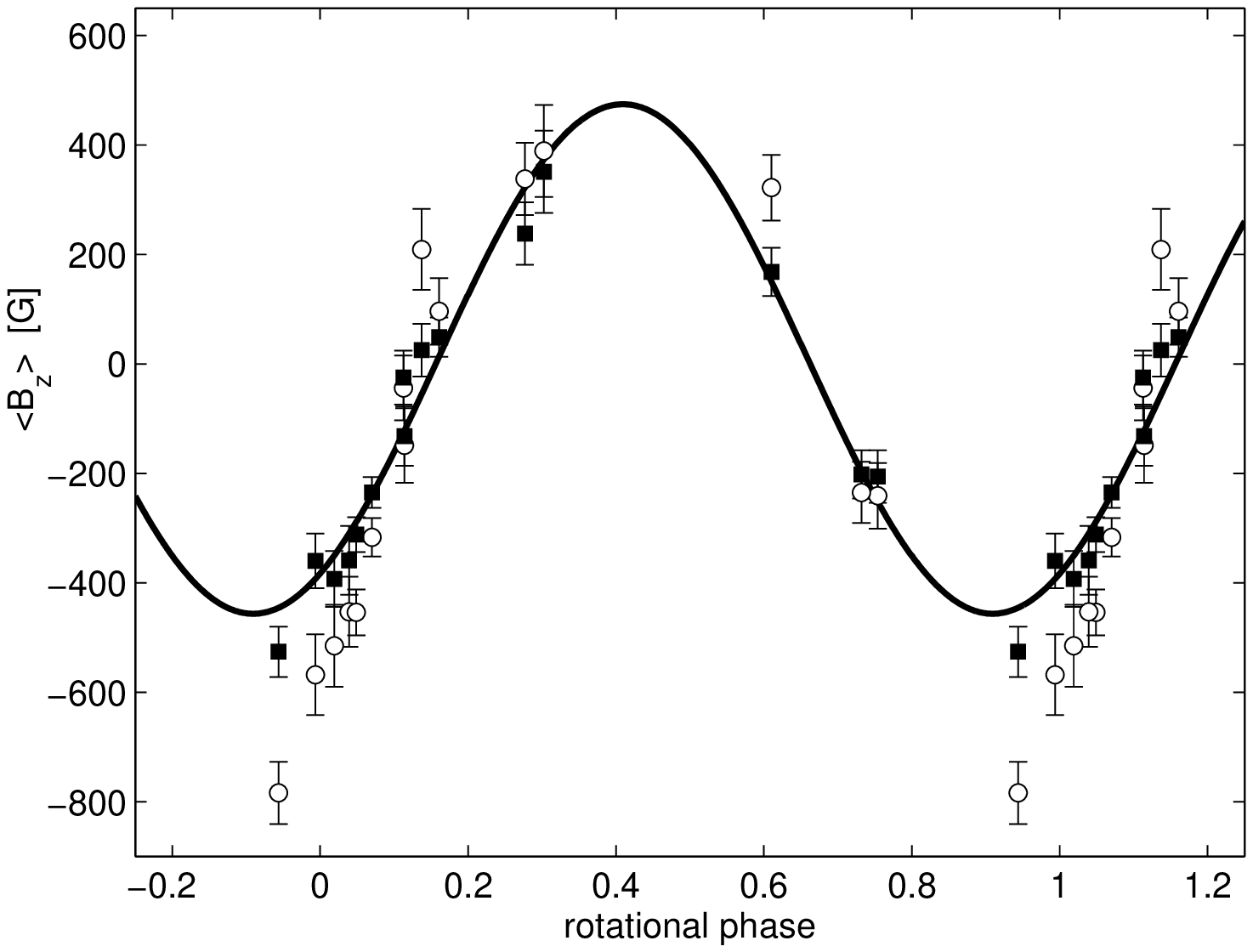}
\includegraphics[width=0.41\columnwidth]{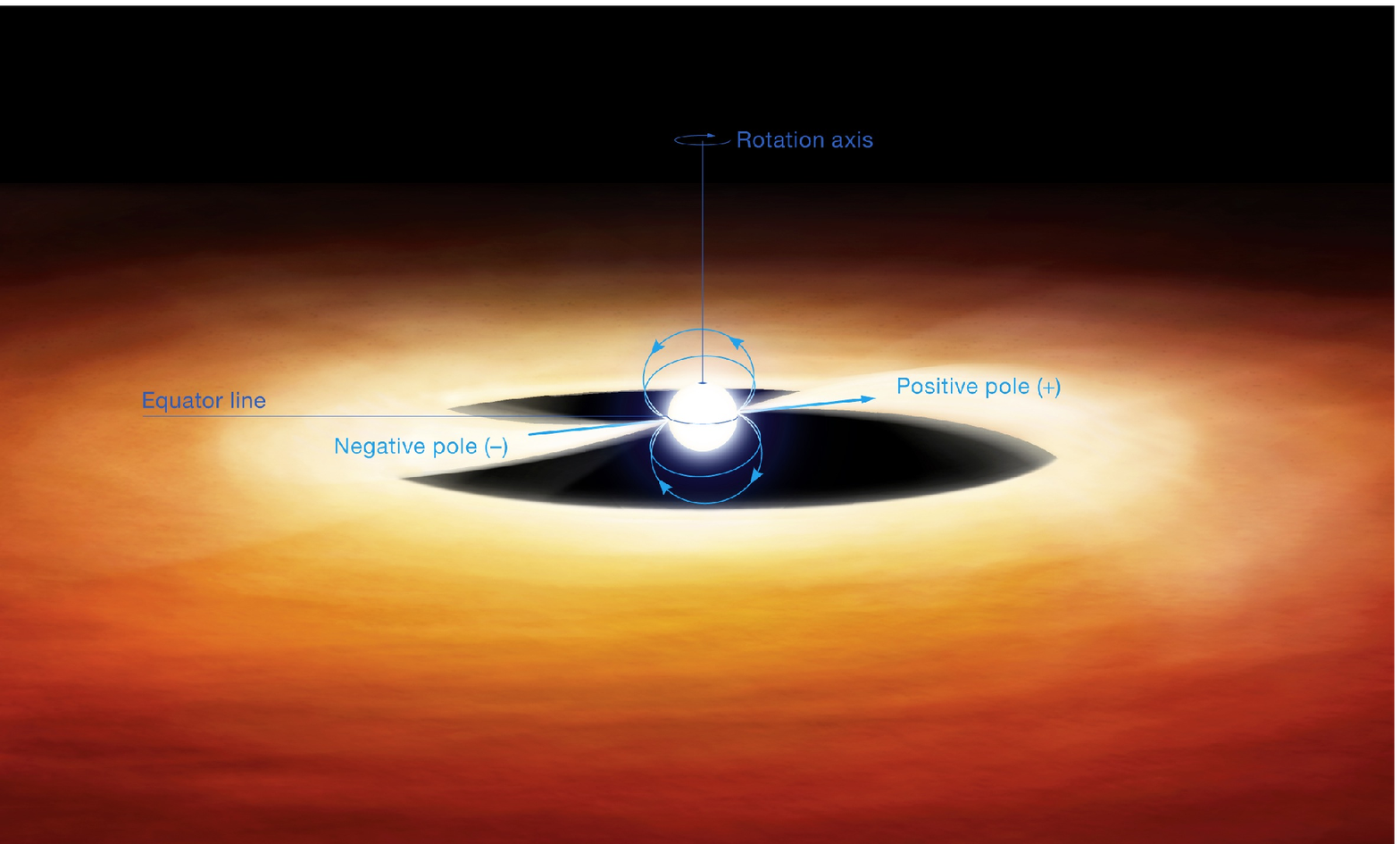}
\caption{
{\em Left:} Phase diagram with the best sinusoidal fit for 
$\left<B_{\rm z}\right>$ measurements of HD\,101412
using all lines (filled squares) and hydrogen lines (open circles;
from \citealt{Hubrig2011}).
{\em Right:}
Artist's impression of the MA in HD\,101412
looking at the magnetic equator
(from \citealt{Schoeller2016}).
}
\label{fig:hd101phasebz}
\end{figure}

The Herbig~Ae star HD\,101412 possesses the strongest magnetic field measured in any Herbig~Ae star so far,
with a surface magnetic field $\left<B\right>$ up to 3.5\,kG.
\citet{Hubrig2011} studied HD\,101412 randomly over its until then unknown
rotation period and found rotational modulation of the longitudinal magnetic field.
From these measurements, they were able to determine several parameters of the star:
the rotation period $P_{\rm rot} = 42.076 \pm 0.017$\,d,
the inclination angle of the rotation axis to the line of sight	$i = 80 \pm 7^{\circ}$,
the obliquity angle, i.e.\ the angle between the rotation axis and the axis of the magnetic dipole, $\beta = 84 \pm 13^{\circ}$
(see Fig.~\ref{fig:hd101phasebz}).

The high value for the magnetic obliquity $\beta = 84\pm13^{\circ}$ challenges theoretical scenarios that
explain MA.
In these scenarios the topology of the channeled accretion critically depends on the magnetic obliquity.
For a large dipole inclination, many magnetic field lines will thread the inner region of the disk matter,
causing strong magnetic braking \citep{Romanova2003}.
This however could explain the long rotation period.

\begin{figure}
\centering
\includegraphics[width=0.36\textwidth]{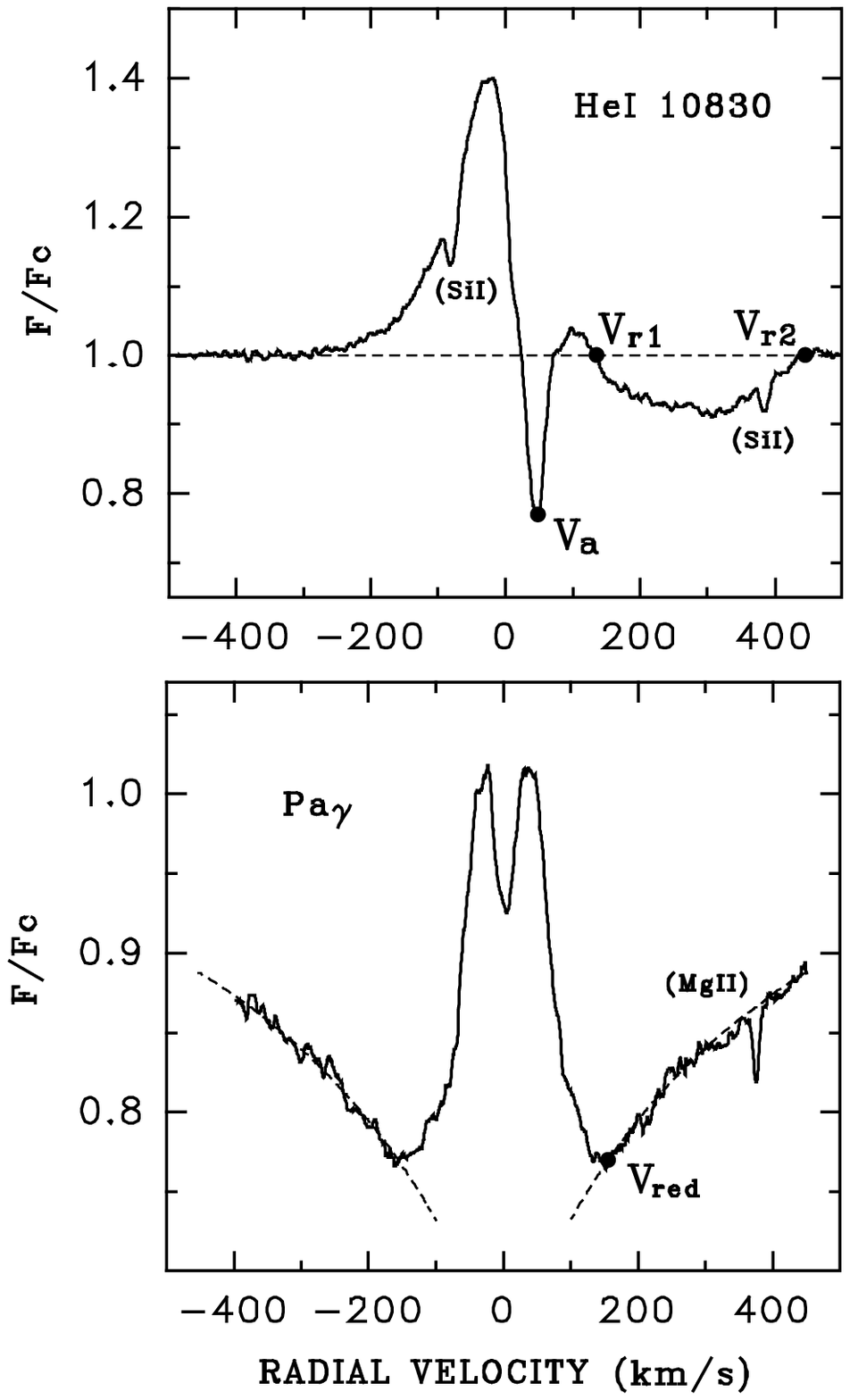}
\includegraphics[width=0.25\textwidth]{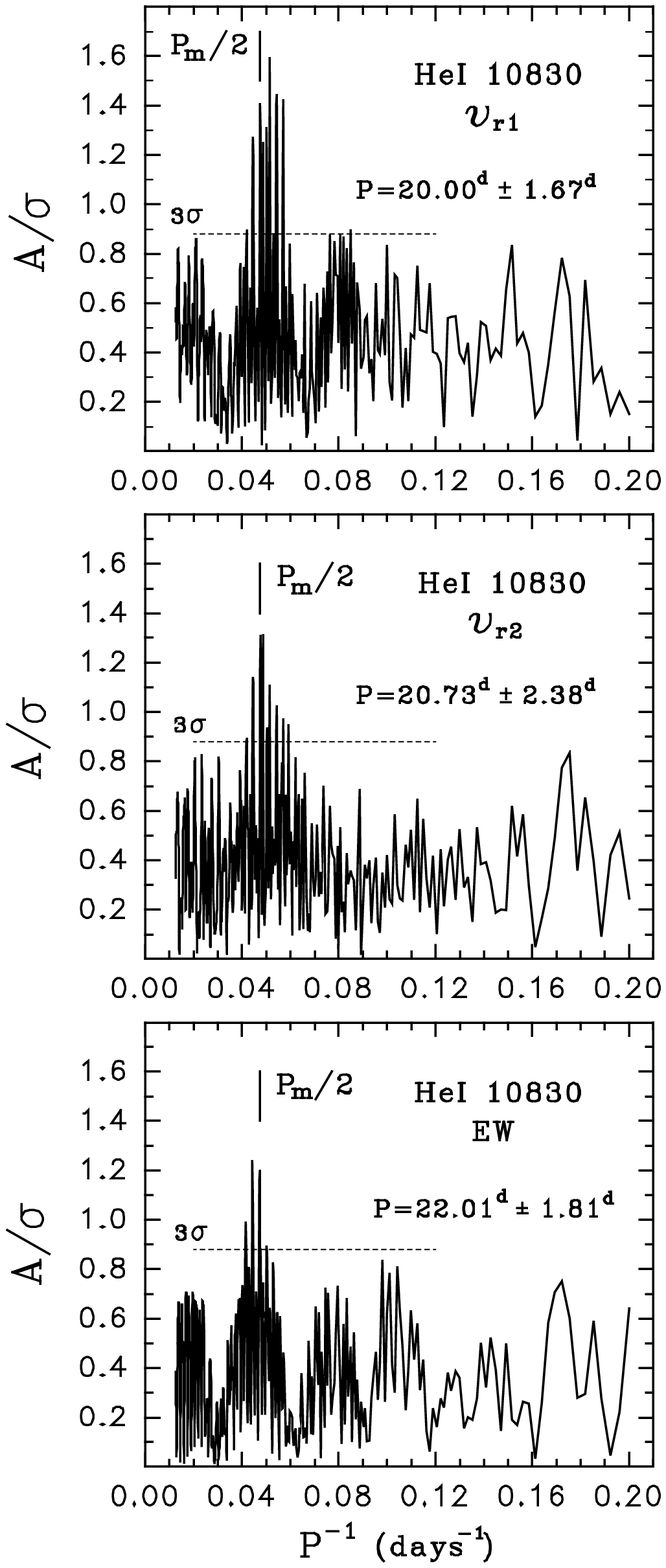}
\includegraphics[width=0.25\textwidth]{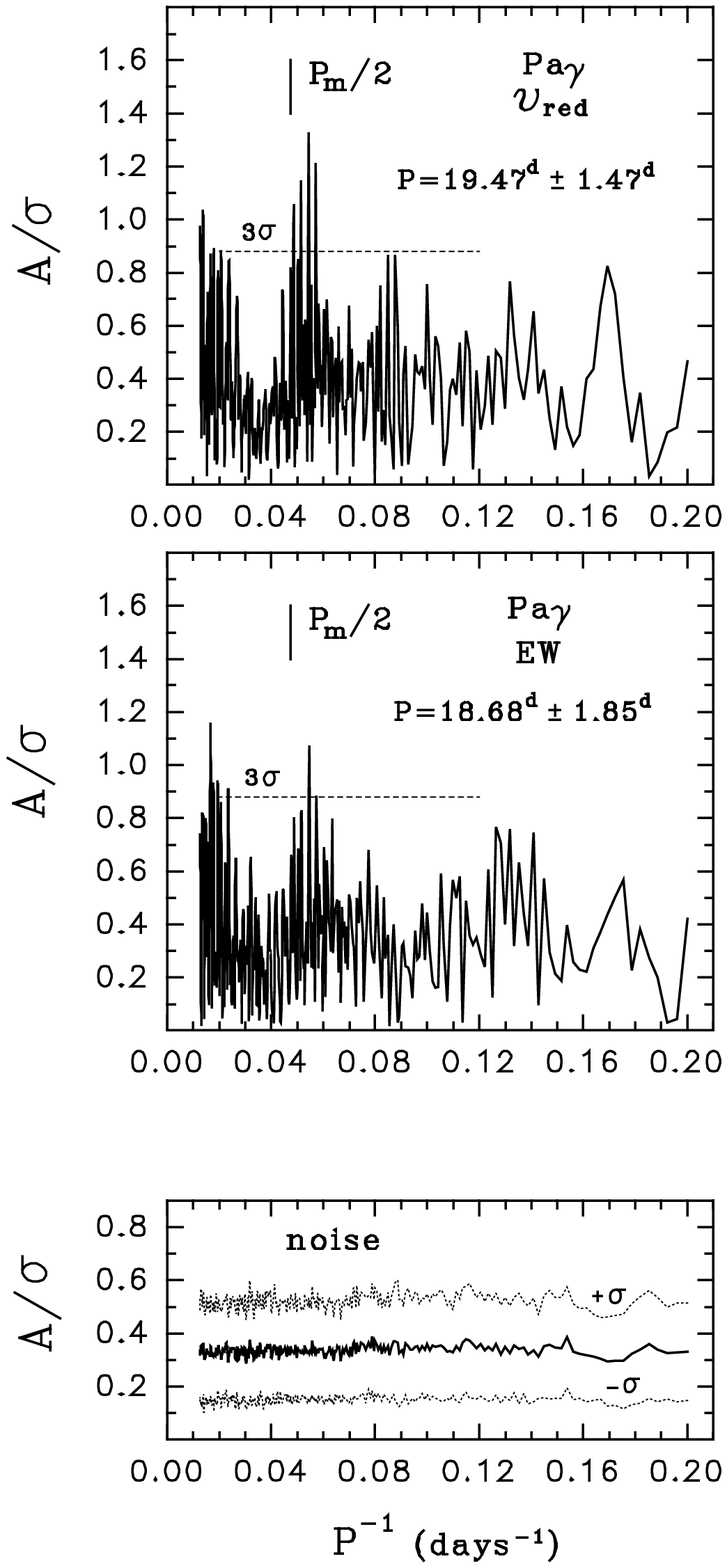}
\caption{
{\em Left:}
Spectral parameters of the \ion{He}{i}~10\,830 and Pa$\gamma$
line profiles used in the quantitative analysis.
{\em Right:}
Different line parameters' $A/\sigma$ periodograms.
Significance levels of 3$\sigma$ are indicated by the dashed lines.
Short vertical lines indicate the value corresponding to half of the magnetic rotation period ($P_{\rm rot}/2 = 21\fd038$).
The detected period values and their errors are given in each plot.
}
\label{fig:hd101period}
\end{figure}

\citet{Schoeller2016} used near-infrared spectroscopic observations of HD\,101412 to
test the magnetospheric character of its accretion.
They analyzed the \ion{He}{i}~10\,830 and Pa$\gamma$ lines
in 30 spectra acquired with the CRIRES and X-shooter spectrographs.
These lines are thought to form in the star's accretion region.
The authors found that the temporal behavior of these diagnostic lines can be
explained by rotational modulation of accreting gas with a period $P = 20\fd53\pm 1\fd68$ (see Fig.~\ref{fig:hd101period}).
The discovery of this period, about half of the magnetic rotation period  $P_{\rm rot} = 42\fd076$,
indicates that the accreted matter falls onto the star in
regions close to the magnetic poles intersecting the line-of-sight twice during the rotation cycle.

\section{HD\,104237}

\begin{figure}
\centering
\includegraphics[width=0.45\textwidth]{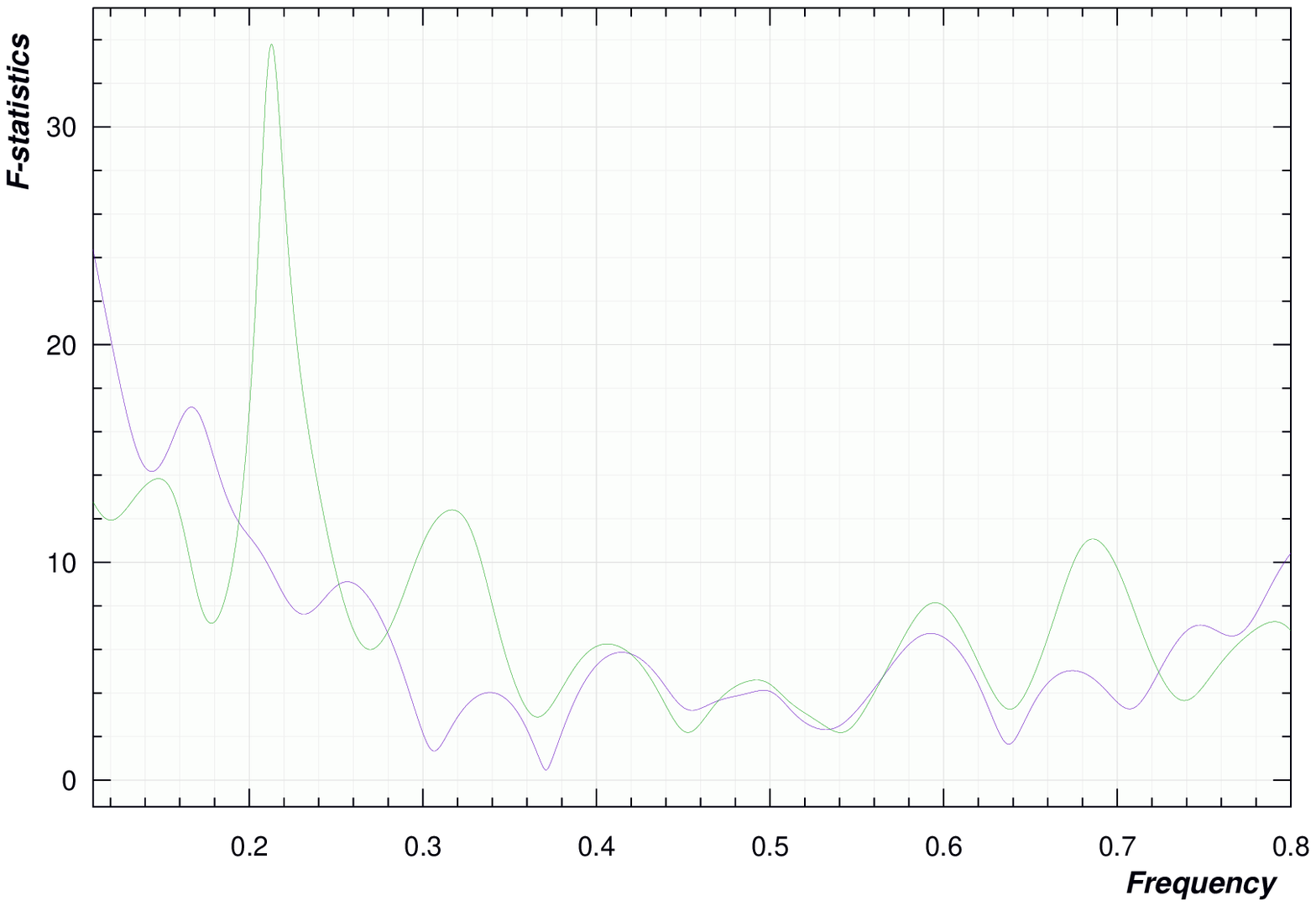}
\includegraphics[width=0.45\textwidth]{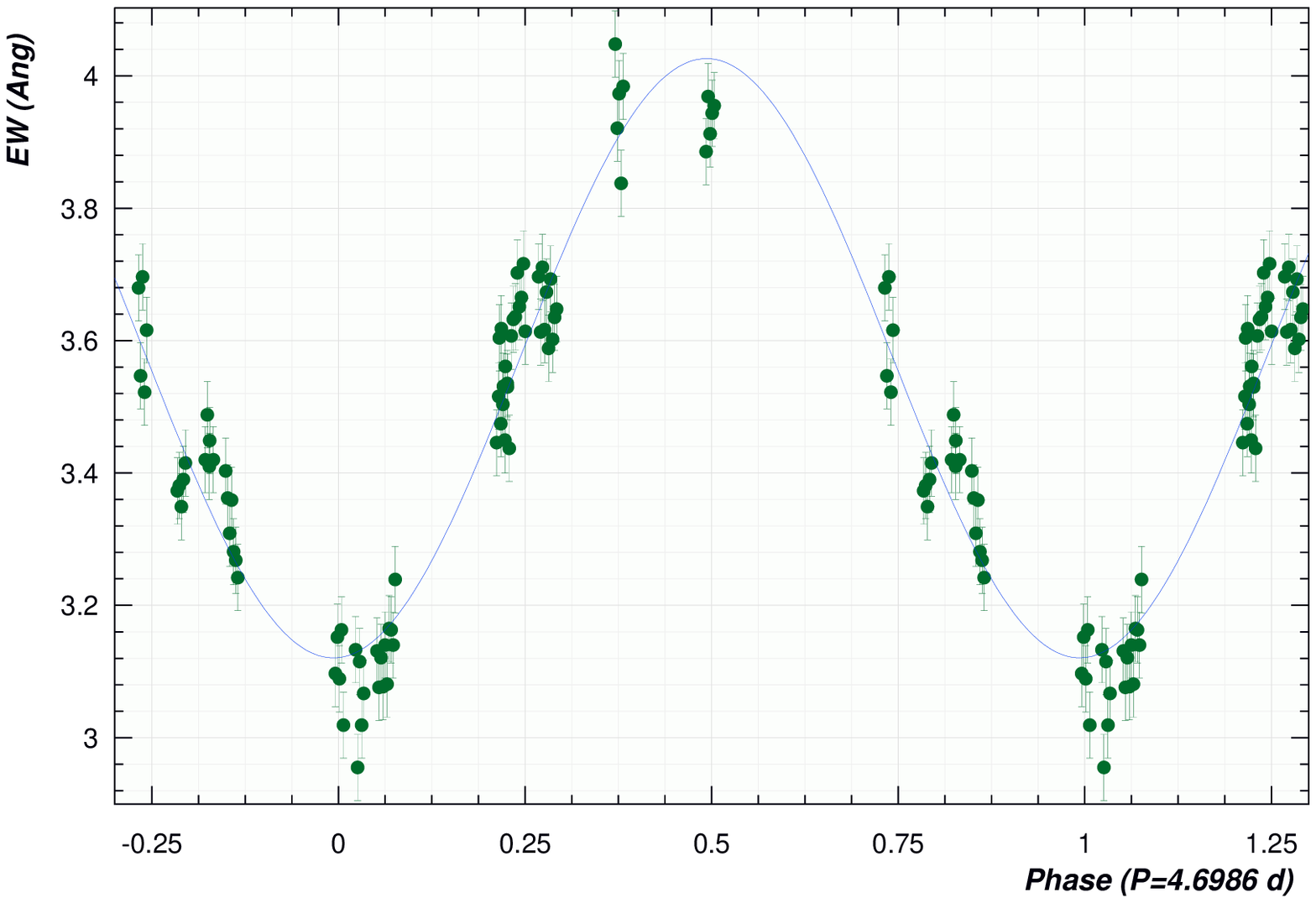}
\caption{
Periodogram (left) and the corresponding phase diagram (right) for HD\,104237 obtained for the $EW$ of
the LSD profiles of 88~HARPS\-pol spectra.
}
\label{fig:hd104period}
\end{figure}

HD\,104237 is a binary system with a Herbig~Ae primary and a T\,Tauri companion.
\citet{Boehm2004} found the orbital period $P_{\rm orb} = 19.86$\,d for the system
and \citet{Boehm2006} the rotation period of the primary $P_{\rm rot,p} = 100\pm5$\,h
from H$\alpha$ measurements.

Using HARPS\-pol spectra from the ESO archive,
J\"arvinen et al.\ (in prep.) determined a highly variable magnetic field,
with values of
$\left<B_{\rm z}\right>=72\pm6$\,G,
$\left<B_{\rm z}\right>=47\pm6$\,G, and
$\left<B_{\rm z}\right>=63\pm6$\,G  for the primary, and
$\left<B_{\rm z}\right>=609\pm27$\,G,
$\left<B_{\rm z}\right>=440\pm23$\,G, and
$\left<B_{\rm z}\right>=124\pm13$\,G for the secondary,
measured over 7.6\,h.

We used eight ISAAC and 13~X-shooter spectra of HD\,104237 from 2013 and 2014,
but our period searches from line parameters
in He\,{\sc i}~5876, He\,{\sc i}~10\,830, and Pa$\gamma$
only resulted in the detection of the non-significant period $P = 5.37$\,d.
However, we were able to show from these data that Pa$\gamma$ originates in the primary.

Computing a least-square-deconvolution (LSD) spectrum for each of 88~archival HARPS\-pol spectra of HD\,104237,
with individual signal-to-noise ratios between 60 and 100, 
and measuring the equivalent width ($EW$) of the resulting line, we found a significant rotation
period for the primary $P_{\rm rot} = 4.7$\,d, which corresponds to 113\,h (see Fig.~\ref{fig:hd104period}).

\section{Summary}

We have demonstrated that it is in principle possible to determine rotation periods in Herbig~Ae/Be stars from accretion tracers.
In practice, these rotation period searches can be hampered by low disk inclinations, binarity, and too sparse time series.
To further improve our knowledge of the magnetic accretion scenario in HAeBes,
we will need a dedicated observing campaign to monitor the magnetic field
in about two dozen HAeBes over their rotation cycle.

\acknowledgements

MAP acknowledges financial support by the Russian Foundation for Basic Research (RFBR) according to research project 18-02-00554
and NAD by RFBR according to research project 18-52-06004.

\bibliography{Schoeller_M}  

\end{document}